\documentclass[aps,12pt,showpacs,superscriptaddress]{revtex4}

\usepackage{graphicx}
\usepackage{mathptmx}

\begin{document}
 
\title{Large-amplitude Electron Oscillations in a Plasma Slab}

\author{L.\ Stenflo} 
\affiliation{Centre for Nonlinear Physics, Department of Physics, 
  Ume{\aa} University, SE--901 87 Ume{\aa}, Sweden}
  
\author{M.\ Marklund}
\affiliation{Centre for Nonlinear Physics, Department of Physics, 
  Ume{\aa} University, SE--901 87 Ume{\aa}, Sweden}
\affiliation{Centre for Fundamental Physics, Rutherford Appleton Laboratory,
Chilton, Didcot, Oxfordshire, UK}

\author{G.\ Brodin} 
\affiliation{Centre for Nonlinear Physics, Department of Physics, 
  Ume{\aa} University, SE--901 87 Ume{\aa}, Sweden}
\affiliation{Centre for Fundamental Physics, Rutherford Appleton Laboratory,
Chilton, Didcot, Oxfordshire, UK}

\author{P.\ K.\ Shukla}
\affiliation{Centre for Nonlinear Physics, Department of Physics, 
  Ume{\aa} University, SE--901 87 Ume{\aa}, Sweden}
\affiliation{Centre for Fundamental Physics, Rutherford Appleton Laboratory,
Chilton, Didcot, Oxfordshire, UK}

\date{\today}

\begin{abstract}
Nonlinear oscillations within a plasma slab are analyzed. Two types of solutions are found, depending on the initial value of the electron density. The first represents regular oscillations within the plasma slab, while the second gives rise to explosive growth at the slab centre or at the edges. The results are discussed. 
\end{abstract}

\maketitle

\section{Introduction}

A few plasma physics problems can be solved exactly (e.g. Akhiezer and Lyubarskii, 1951; Dawson, 1959; Davidson, 1972; Shivamoggi, 1988; Stenflo and Yu, 1997). The plasma disturbances have in those cases been allowed to be so large that no expansion techniques are applicable. The new exact solutions can be useful in interpreting observed large amplitude wave phenomena in detail, as well as for verifying new approximation schemes and numerical methods in the study of nonlinear effects.
We shall in the present paper reconsider one of the simplest possible situations, namely that of one-dimensional ($\partial_y = \partial_z = 0$) oscillations in a cold plasma slab ($-d \leq x \leq d$) where the electrons are mobile, whereas the ions are immobile and form a fixed background with the constant density $n_0$. The special case where the electron density $n$ is only a function of time and where the cold electron fluid velocity $v(x,t)$ satisfies the boundary conditions $v(\pm d, t) = 0$ has been solved previously (Aliev and Stenflo, 1994; Stenflo, 1996). Here we shall use the same boundary condition but consider a more general and useful class of solutions for $n(x,t)$.

\section{Basic equations}

Our governing equations are
\begin{eqnarray}
  && \partial_t n + \partial_x(nv) = 0 , \label{eq:cont} \\
  && \partial_tv + v\partial_xv = -(e/m)E , \label{eq:mom} \\
  && \partial_xE = (e/\epsilon_0)(n_0 - n) , \label{eq:poisson}
\end{eqnarray}
where $E$ is the electric field, $e$ is the magnitude of the electron charge, $m$ is the electron mass, and $\epsilon_0$ is the vacuum dielectric constant. By eliminating the electric field in (\ref{eq:mom}) and (\ref{eq:poisson}) we immediately find the equation
\begin{equation}\label{eq:elim}
  \partial_x(\partial_tv + v\partial_xv) = -\omega_p^2(1 - n/n_0) ,
\end{equation}
where $\omega_p^2 = n_0e^2/\epsilon_0m$.

We shall below study a particular solution where $n(x,t)$ and $v(x,t)$ satisfy the system of two coupled equations (\ref{eq:cont}) and (\ref{eq:elim}), and where $v(\pm d, t)  = 0$. Our boundary condition means that the electrons are always contained within the slab, and that $\int_{-d}^dn(x,t)\,dx$ accordingly is a constant ($ = 2d\,n_0$). By means of straightforward calculations it is then easy to verify that a solution of (\ref{eq:cont}) and (\ref{eq:elim}) is 
\begin{equation}
  n(x,t) = \frac{n_0}{1 + \kappa'(y)\cos(\omega_pt)} , \label{eq:density}
\end{equation}
and
\begin{equation}
  v(x,t) = -d\,\omega_p \kappa(y)\sin(\omega_pt), \label{eq:velocity}
\end{equation}
where $y(x,t)$ is given by the implicit relation 
\begin{equation}\label{eq:x}
  x = d\,[y + \kappa(y)\cos(\omega_pt)] , 
\end{equation}  
and the prime denotes differentiation with respect to $y$. Here 
$\kappa(y)$ is an arbitrary function satisfying $\kappa(\pm 1) = 0$. We shall in the present paper  choose the simple function $\kappa(y) = \alpha y(1 - y^2)$, where $\alpha$ is a dimensionless constant. 
With this choice of $\kappa$ we note that (\ref{eq:x}) is a third order polynomial in $y$, and that an analytical expression $y(x,t)$ thus can be presented by solving 
\begin{equation}\label{eq:x-part}
  x = d\,[y + \alpha y(1 - y^2)\cos(\omega_pt)] .
\end{equation}
However, here we do not give the explicit expression for $y(x,t)$. Other choices of $\kappa$ lead to more elaborate expressions for $y(x,t)$ and will therefore not be considered. 
Corresponding to (\ref{eq:x-part}), (\ref{eq:density}) and (\ref{eq:velocity}) now reduce to
\begin{equation}
  n(x,t) = \frac{n_0}{1 + \alpha(1 - 3y^2)\cos(\omega_pt)} , \label{eq:density2} 
\end{equation}
and
\begin{equation}
  v(x,t) = -d\,\omega_p \alpha y(1 - y^2)\sin(\omega_pt). \label{eq:velocity2}
\end{equation}

We note that $n(\pm d,0) = n_0/(1 - 2\alpha)$.  
With the normalization $n \rightarrow n/n_0$, $v \rightarrow v/d\,\omega_p$, $x \rightarrow x/d$, $t \rightarrow \omega_pt$, and $d \rightarrow 1$ we plot the density and velocity profiles for different times. One can then verify that the total number of electrons is conserved during the oscillations. 
First, when $0 < \alpha < 1/2$ we show (in the left panel of Fig.\ 1) $n(x,t)$ for different normalized times $t$, using $\alpha = -0.4$. We note the symmetric and oscillating character of $n$. The corresponding velocity profile $v(x,t)$ can be seen in the right panel of Fig.\ 1. For these regular oscillations, the sign of $\alpha$ characterizes the initial value of the electron density $n(x,0)$: when $\alpha >0$ we start with higher edge density at $x = \pm d$, while for $\alpha < 0$ the density has an initial spatial maximum at $x =0$. 

\begin{figure*}
\includegraphics[width=.48\textwidth]{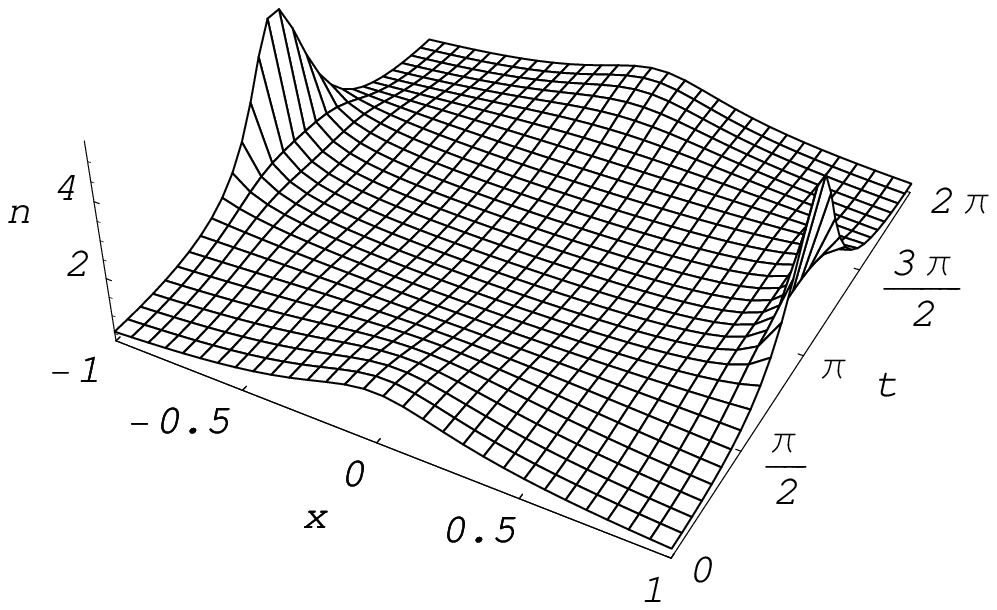}
\includegraphics[width=.48\textwidth]{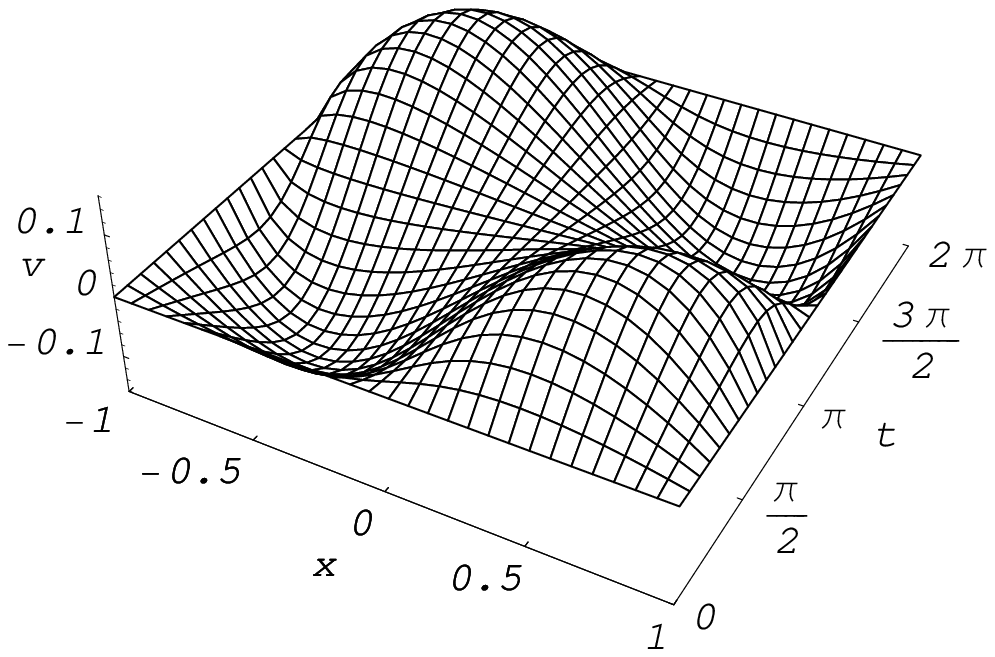}
\caption{\it Nonlinear periodic and symmetric density oscillations. To the left the normalized density $n$ is plotted and to the right the normalized velocity $v$ is plotted, both as functions of the normalized length $x$ for different normalized times $t$, with $\alpha = -0.4$. 
At $t = 0$ the maximum density can be seen at $x = 0$. When $\alpha > 0$ we have the same dynamics shifted by a quarter of a period.}
\end{figure*}

The denominator in (\ref{eq:density2}) can approach zero at a finite time $t_0$,  resulting in explosive growth. Due to the symmetric character of the density distribution the explosion for $t = t_0$ will either be at (a) the edges of the slab ($x = \pm d$, i.e.\ $y = \pm 1$) or (b) at the centre of the slab ($x = 0$, i.e.\ $y = 0$). In the case (a), the explosion time $t_0$ is given by $\cos(\omega_pt_0) = 1/2\alpha$ and $|\alpha| > 1/2$, while in case (b) the explosion time is found from $\cos(\omega_pt_0) = -1/\alpha$ and $|\alpha| > 1$. Thus, in case (b) we can have explosive growth both at the edges and at the centre, depending on the initial conditions, while this it not necessarily true for case (a). The explosive instability occurs in the wave-breaking regime, and our cold plasma model thus breaks down at this stage (see Fig.\ 2).
A change of sign of $\alpha$ shifts the density distribution by half a period. Note however that $\alpha > 0$ gives rise to unphysical initial conditions, i.e.\ $n(\pm d,0) < 0$.

\begin{figure*}
\includegraphics[width=.48\textwidth]{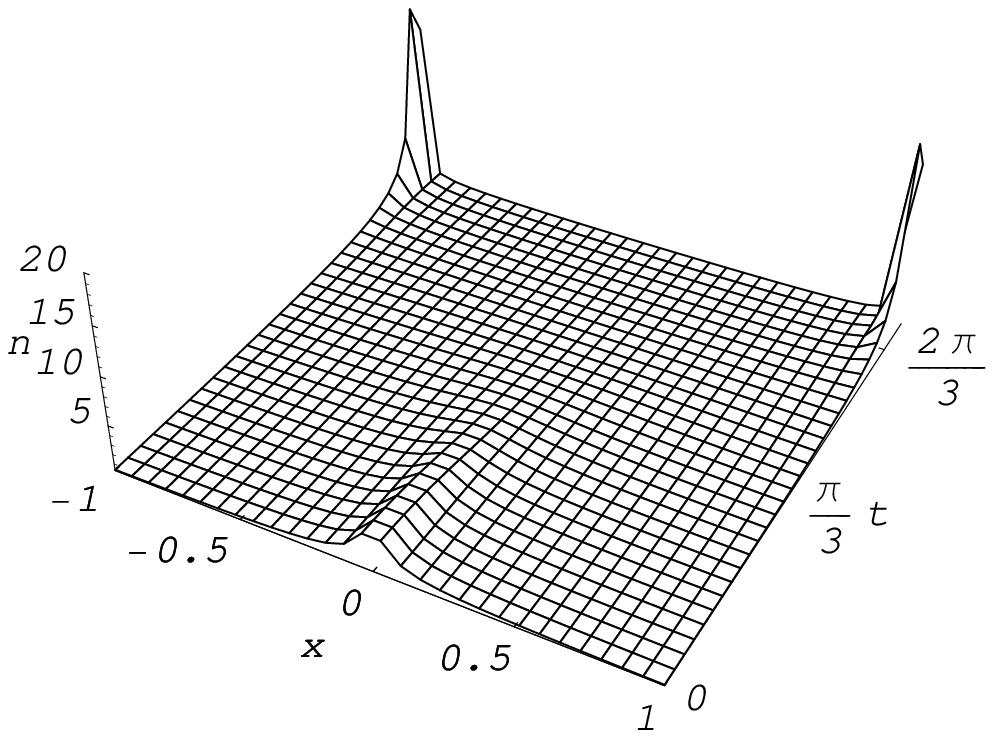}
\includegraphics[width=.48\textwidth]{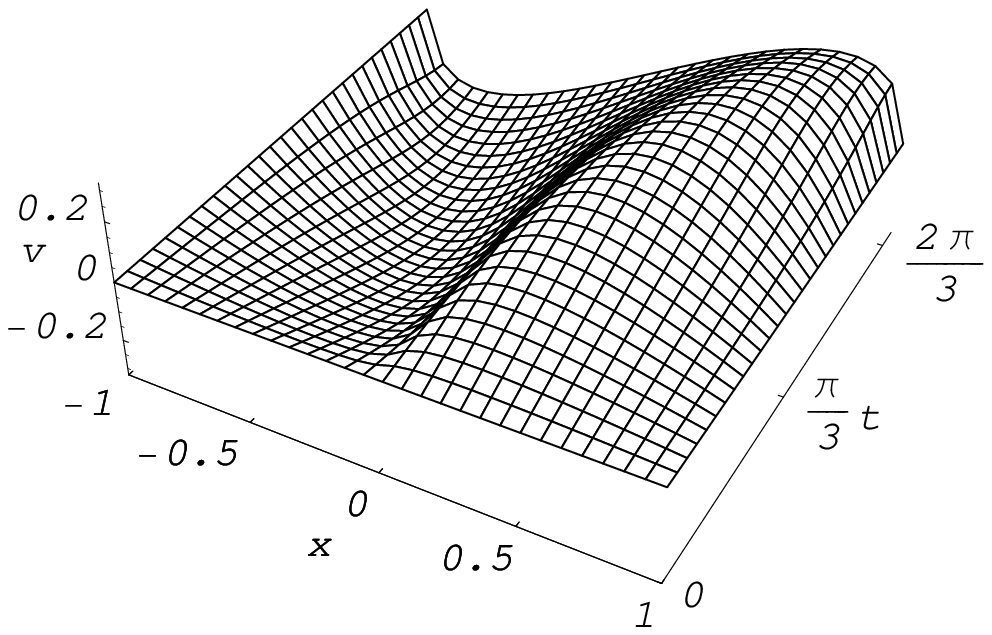}
\caption{\it Explosive instability and wave breaking. To the left the normalized density $n$ is plotted and to the right the normalized velocity $v$ is plotted, both as functions of the normalized length $x$ for different normalized times $t$, with $\alpha = -0.8$. 
At $t = 0$ the density is focused but finite at $x = 0$. As the system evolves we can see wave breaking and density spikes developing at the edges of the slab. Here the case $\alpha > 0$ is no longer associated with a simple phase shift, since wave breaking then occurs at $t = 0$.}
\end{figure*}

\section{Conclusion}

In the present paper we have generalized the work of Aliev and Stenflo (1994) to allow for a class of solutions where the electron density $n(x,t)$ is an explicit, and comparatively simple, function of $x$. We have presented figures which show the behavior of the solutions for different initial values $n(x,0)$. From those solutions we can very clearly see that $n(x,t)$ has an oscillating behavior when $0 < |\alpha| < 1/2$, while if $|\alpha| > 1/2$ we have explosive growth, i.e.\ $n$ grows like $1/(t - t_0)^2$ where the explosion time $t_0$ is of the order $1/\alpha\omega_p$. 
We think that the present solution can be generalized to cold many-component plasmas (Amiranashvili et al., 2002), to plasmas where $n_0$ is a function of $x$ (Stenflo and Gradov, 1990; Karimov, 2002), to cylindrical plasmas (Stenflo and Yu, 1997; Karimov, 2005), to Pierce beam plasma systems (Matsumoto, Yokohama, and Summers, 1996), 
as well as to other kind of plasmas (Vladimirov, Yu, and Tsytovich, 1994). Alternative solutions are of course also possible (Polyakov, 1995).

Our governing equations are easily generalized to the relativistic case by replacing Eq.\ (\ref{eq:mom}) by $\partial_tp + v\partial_xp = -eE$, where $p = mv/\sqrt{1 - v^2/c^2}$, $c$ is the velocity of light, and $m$ is the electron rest mass. At present we have however not been able to generalize (\ref{eq:density}) and (\ref{eq:velocity}) to the relativistic case, and we have therefore limited our presentation to the regime $d\omega_p \sim v \ll c$. The opposite limit is of course also of much interest, but it has to be delegated to future numerical work.

\acknowledgments

This research was partially supported by the Swedish Research Council.

\end{document}